\documentclass[aps,pra,twocolumn,,superscriptaddress,floatfix]{revtex4-1}
\usepackage[svgnames]{xcolor}
\definecolor{darkred}{rgb}{0.6, 0, 0}
\definecolor{darkgreen}{rgb}{0, 0.5, 0}
\usepackage[colorlinks=true,urlcolor = purple,linkcolor = darkgreen, citecolor = magenta, bookmarks=false]{hyperref}
\usepackage{amsfonts,amsmath,amssymb,amsthm}
\usepackage{graphicx}
\usepackage{dsfont}
\usepackage[normalem]{ulem}
\usepackage{enumerate}
\usepackage{natbib}

\usepackage{tcolorbox}
\usepackage{tabularx}
\usepackage{array}
\usepackage{colortbl}
\tcbuselibrary{skins}

\swapnumbers

\def\ii{{\rm i}}
\newcommand{\dd}{{\rm d}}

\usepackage{tikz}
\usepackage{pgfplots}
\usetikzlibrary{arrows,topaths,shapes.geometric,shapes.misc,patterns,positioning,shadows,decorations.markings,
decorations.pathreplacing,calc, trees, positioning, arrows, shapes, shapes.multipart, shadows, matrix, decorations.pathreplacing, decorations.pathmorphing}
\tikzset{->-/.style={decoration={
  markings,
  mark=at position #1 with {\arrow[scale=1.5]{latex}}},postaction={decorate}}}

\usepackage{pgfplots}
\pgfplotsset{compat=1.9}

\begin{document}

\title{Absence of Normal Fluctuations in an Integrable Magnet
}

\author{\v{Z}iga Krajnik}
\affiliation{Faculty for Mathematics and Physics,
University of Ljubljana, Jadranska ulica 19, 1000 Ljubljana, Slovenia}

\author{Enej Ilievski}
\affiliation{Faculty for Mathematics and Physics,
University of Ljubljana, Jadranska ulica 19, 1000 Ljubljana, Slovenia}

\author{Toma\v{z} Prosen}
\affiliation{Faculty for Mathematics and Physics,
University of Ljubljana, Jadranska ulica 19, 1000 Ljubljana, Slovenia}

\date{\today}

%%%%%%%%%%%%%%%%%%%%%%%%%%%%%%%%%%%%%%%%%
%%%%%%%%%%%%%%%%%%%%%%%%%%%%%%%%%%%%%%%%%
\begin{abstract}
We investigate dynamical fluctuations of transferred magnetization in the  one-dimensional lattice Landau--Lifshitz magnet with uniaxial anisotropy, representing an emblematic model of interacting spins.
We demonstrate that the structure of fluctuations in thermal equilibrium depends radically on the characteristic dynamical scale.
In the ballistic regime, typical fluctuations are found to follow a normal distribution and scaled cumulants are finite. In stark contrast, on the diffusive and superdiffusive timescales, relevant respectively for the easy-axis and isotropic magnet at vanishing total magnetization, typical fluctuations are no longer Gaussian and, remarkably, scaled cumulants are divergent.
The observed anomalous features disappear upon breaking integrability, suggesting that the absence of normal fluctuations is
intimately tied to the presence of soliton modes.
In a nonequilibrium setting of the isotropic magnet with weakly polarized step-profile initial state we find a slow drift of dynamical exponent from the superdiffusive towards the diffusive value.
\end{abstract}


\maketitle

{\em Introduction.}---Explaining how phenomenological laws of physics emerge on macroscopic scales from reversible microscopic dynamics underneath presents a formidable task. The challenge only grows in many-body interacting systems, both in and out of equilibrium,
where analytic results without resorting to assumptions or uncontrolled approximations are rarely available. This explains, at least in part, 
the perpetual fascination with exactly solvable models and stimulates our quest for non-trivial exact results.

It has long been known that one-dimensional systems occupy a very special place in this regard, hosting a wide range of unorthodox phenomena such as lack of conventional thermalization \cite{CCR11,VR16,EF_review,Caux_QA,PhysRevLett.115.157201,QLreview}, anomalous transport behavior \cite{transport_review,GHD_review,superdiffusion_review} and unconventional entanglement properties \cite{Alba18,Calabrese20}.
Integrable models defy ordinary hydrodynamics \cite{DeNardis18,Gopalakrishnan18,DYC18,DDKY19,Bastianello19} due to ballistically propagating quasiparticles stabilized by infinitely many conservation laws. This readily explains why many of their dynamical properties
are markedly different from generic (i.e. ergodic) systems, such as non-zero finite-temperature Drude weights \cite{CZP95,Zotos99,Prosen11,IN_Drude,DS17,IN17} or superdiffusive spin transport in models with nonabelian symmetries
that has sparked great theoretical interest
both in quantum  \cite{Marko11,Bojan,Ljubotina17,Ilievski18,Ljubotina19,GV19,NMKI19,DupontMoore19,Vir20,superuniversality} and classical \cite{Bojan,Das19_KPZ,KP20,MatrixModels} integrable models, see Ref.~\cite{superdiffusion_review} for a review.
Understanding these aspects goes beyond just theoretical interest. Experimental techniques with
cold atoms have now finally advanced to the point to enable the fabrication of various low-dimensional paradigms \cite{Langen15,Schemmer19,Weiner20,Jepsen20,Malvania21,Scheie21,Joshi21,Bloch_KPZ}, thereby offering a great opportunity
to directly probe many different facets of nonequilibrium phenomena.

A more refined information about dynamical processes, extending beyond hydrodynamics, can be inferred by inspecting the structure of fluctuating macroscopic quantities.
In this respect, large deviation (LD) theory \cite{Touchette_LDT,Esposito_review,JuanP18} has cemented itself as a versatile theoretical apparatus designed to quantify the probability of rare events. It is quite remarkable that in certain scenarios the large deviation rate function can be deduced analytically, including the Levitov--Lesovik formula~\cite{LL94,LLL96},
free fermionic systems~\cite{Moriya19,Gamayun20} and field theories~\cite{Yoshimura18,DLSB15}, 
noninteracting~\cite{Znidaric14_LD,Znidaric14_LD_diff} and interacting~\cite{Buca14,Matthieu19} systems with dissipative boundary
driving, conformal field theories \cite{BD15,BD16}, in conjunction with a body of exact results
from the domain of classical stochastic gases \cite{deGier05,GM06,Derrida07,Derrida09,Lazarescu13}.
While in classical diffusive systems the rate function can be in principle deduced within the framework of macroscopic fluctuation theory (MFT) \cite{Bertini02,MFT}, the resulting equations typically prove difficult to handle. 
A general LD theory for classical and quantum integrable systems on ballistic scales has been developed in Refs.~\cite{MBHD20,DoyonMyers20,PerfettoDoyon21}.

In spite of tremendous progress, it nonetheless appears that in \emph{deterministic} (Hamiltonian) many-body systems of \emph{interacting} degrees of freedom there are, except for a numerical survey in non-integrable anharmonic chains \cite{MendlSpohn}, no explicit results concerning the nature of typical or large fluctuations, especially on subballistic scales. This motivates the study
of integrable systems, which are promising candidates to reveal novel unorthodox features due to their distinct non-ergodic properties.
Additional inspiration comes from an earlier study \cite{Znidaric_anomalous} of the anisotropic quantum Heisenberg chain driven out of equilibrium by means of Lindbladian baths that hints at anomalous scaling of higher cumulants in the gapped
(i.e. diffusive) phase of the model (albeit for moderately small system sizes), suggesting that despite a well-defined
diffusion constant, the gapped Heisenberg chain may not be an ordinary diffusive conductor.
Efficient simulations of quantum dynamics are unavoidably hampered by a rapid increase of entanglement which often precludes a reliable extraction of asymptotic scaling laws. This shortcoming motivates the study of \emph{classical} integrable models where this is no longer
a concern and much longer simulation times are accessible.

In this paper, we examine fluctuations of spin current over a finite time interval in a thermodynamic ensemble of interacting classical spins evolving under a deterministic integrable dynamics. In our simulations we take full advantage of a symplectic integrator
developed in Ref.~\cite{LLMM} which \emph{exactly} preserves integrability.
\begin{widetext}

\begin{figure}[t]
\centering
\includegraphics[width=\columnwidth]{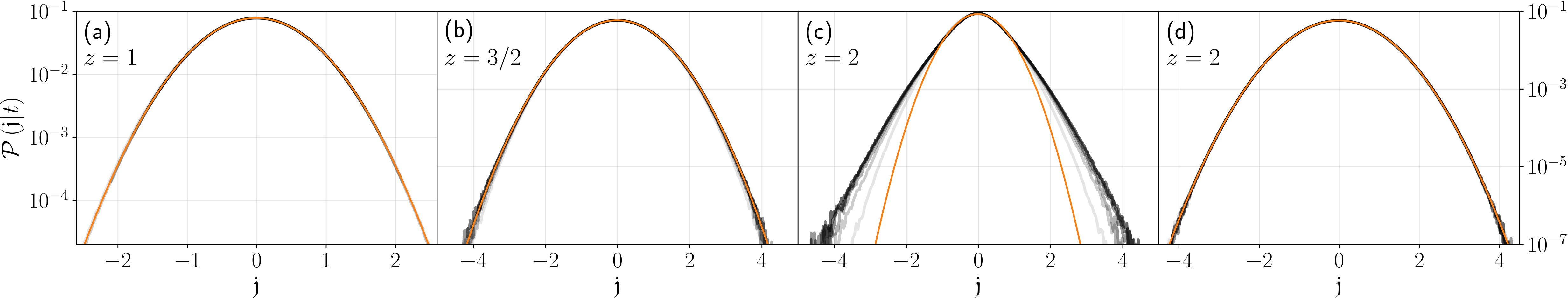}
\caption{Convergence of distributions $\mathcal{P}(\mathfrak{j}|t)$ for various dynamical regimes: (a) {\em integrable} easy-plane regime with ballistic exponent $z=1$, 
(b) {\em integrable} isotropic model with 
superdiffusive exponent $z = 3/2$ and (c) {\em integrable} easy-axis regime with diffusive exponent $z=2$. For comparison, (d) is a {\em non-integrable} isotropic trotterization with $z = 2$.
Best-fit Gaussian distributions are depicted by full orange lines. Time increases from gray $(t=40)$ to black $(t=500)$. Simulation parameters: timestep $\tau = 1$, length $L = 2^{10}$, anisotropy $\varrho = \gamma=1$, $N_{\textrm{avg}} = 3 \times 10^{5}$.
The models and conventions are given in \cite{suppmat}.}
\label{fig:figure1}
\end{figure}
\end{widetext}
Strikingly, in the diffusive and superdiffusive dynamics regimes we encounter hitherto undisclosed anomalous fluctuations and divergent scaled cumulants.

{\em Fluctuations on typical scale.}--- Our main objective is to characterize the dynamics of magnetization
in a one-dimensional classical spin system governed by a \emph{deterministic} equation of motion for the spin field
${\bf S}\equiv (S^{1},S^{2},S^{3})^{\rm T}$ (subject to constraint $|{\bf S}|=1$).
We are specifically interested in extended (i.e. thermodynamic) homogeneous systems of interacting spins
in which the third component of total magnetization is a globally conserved charge, $Q=\int \dd x\,S^{3}(x)$,
satisfying a local continuity equation $\partial_{t}S^{3}(x,t)+\partial_{x}j(x,t)=0$.

In this work, we aim to characterize the fluctuations of the time-integrated spin current density passing through the origin
in a finite interval of length $t$,
\begin{equation}
\mathfrak{J}(t) = \int^{t}_{0}\dd t'\,j(0,t').
\label{eqn:integrated_current}
\end{equation}
Equation \eqref{eqn:integrated_current} represents the net transferred magnetization between two semi-infinite regions of the system
that can be regarded as a fluctuating macroscopic dynamical variable, and the main scope of this work is to examine its statistical properties in thermal equilibrium. While on average there is no transferred charge, $\langle \mathfrak{J}(t)\rangle = 0$, the variance of $\mathfrak{J}(t)$ at large times times grows algebraically
with an equilibrium \emph{dynamical exponent} $z$,
\begin{equation}
\left\langle [\mathfrak{J}(t)]^2\right\rangle^{c} \sim t^{1/z},
\label{z_def}
\end{equation}
where $\langle \bullet \rangle^{c}$ denotes the connected part of the $n$-point correlation in thermal equilibrium.
For simplicity, we shall subsequently compute averages with respect to an unbiased stationary measure, representing the high-temperature limit of the canonical Gibbs ensemble.

Notice that \emph{typical} fluctuations of $\mathfrak{J}(t)$ are of the order $\mathcal{O}(t^{1/2z})$.
In order to quantify them, we introduce the dynamical distribution $\mathcal{P}(\mathfrak{j}|t)$ of the \emph{scaled integrated current density}
\begin{equation}
\mathfrak{j}(t)
 \equiv t^{-1/2z}\mathfrak{J}(t),
\end{equation}
and subsequently determine the \emph{stationary} probability distribution that may emerge at large times,
$\mathcal{P}(\mathfrak{j}) = \lim_{t\to \infty} \mathcal{P}(\mathfrak{j}|t)$, normalized as $\int \mathcal{P}(\mathfrak{j}|t)\dd \mathfrak{j}=1$.
%\begin{equation}
%\mathcal{P}(\mathfrak{j}) = \lim_{t\to \infty} t^{1/2z}P(\mathfrak{j};t),\qquad \int \mathcal{P}(\mathfrak{j})\dd \mathfrak{j}=1.
%\label{eqn:stationary_distribution}
%\end{equation}
We shall characterize it by its cumulants $\kappa_{n}=\lim_{t\to \infty} \kappa_n(t)$,
\begin{equation}
\kappa_n(t) \equiv \left\langle [\mathfrak{j}(t)]^n\right\rangle^c =
t^{-n/2z}  \left\langle [\mathfrak{J}(t)]^n\right\rangle^c.
\end{equation}
By the time-reversal symmetry of an equilibrium state $\mathcal{P}(\mathfrak{j})$ is symmetric and hence $\kappa_{2n+1}=0$
for all $n\in \mathbb{N}$.
If all $\kappa_{n\neq 2}=0$, then $\mathcal{P}(\mathfrak{j})$ takes the form of a Gaussian and typical fluctuations are said to be \emph{normal}. For rapidly decaying temporal correlations of local currents, 
this property, for $z=1$, is indeed guaranteed by the central limit theorem, as eg. in rule 54 dynamics~\cite{Klobas}.
In Hamiltonian systems, temporal correlations of currents of conserved charges are invariably present
and very little is known about their clustering properties, therefore no general conclusions about the dynamical exponent and Gaussianity of $\mathfrak J(t)$ can be drawn. 
However, in non-integrable (chaotic) systems having the hydrodynamic mode with zero velocity, one may expect spatiotemporal correlations to follow diffusive phenomenology, implying dynamical exponent $z=2$ and Gaussian fluctuations.
%According to empirical evidence at least,
%they generically decay sufficiently rapidly to effectively restore `Gaussianity' at late times.
%Integrable models may be an exception in this regard as their dynamical correlations are known to be radically different from those of ergodic systems.
%{\color{red}TP: prejsnjih stavkov ne razumem, to velja le v balisticnem rezimu, ali???}

\begin{figure}[b]
\centering
\includegraphics[width=\columnwidth]{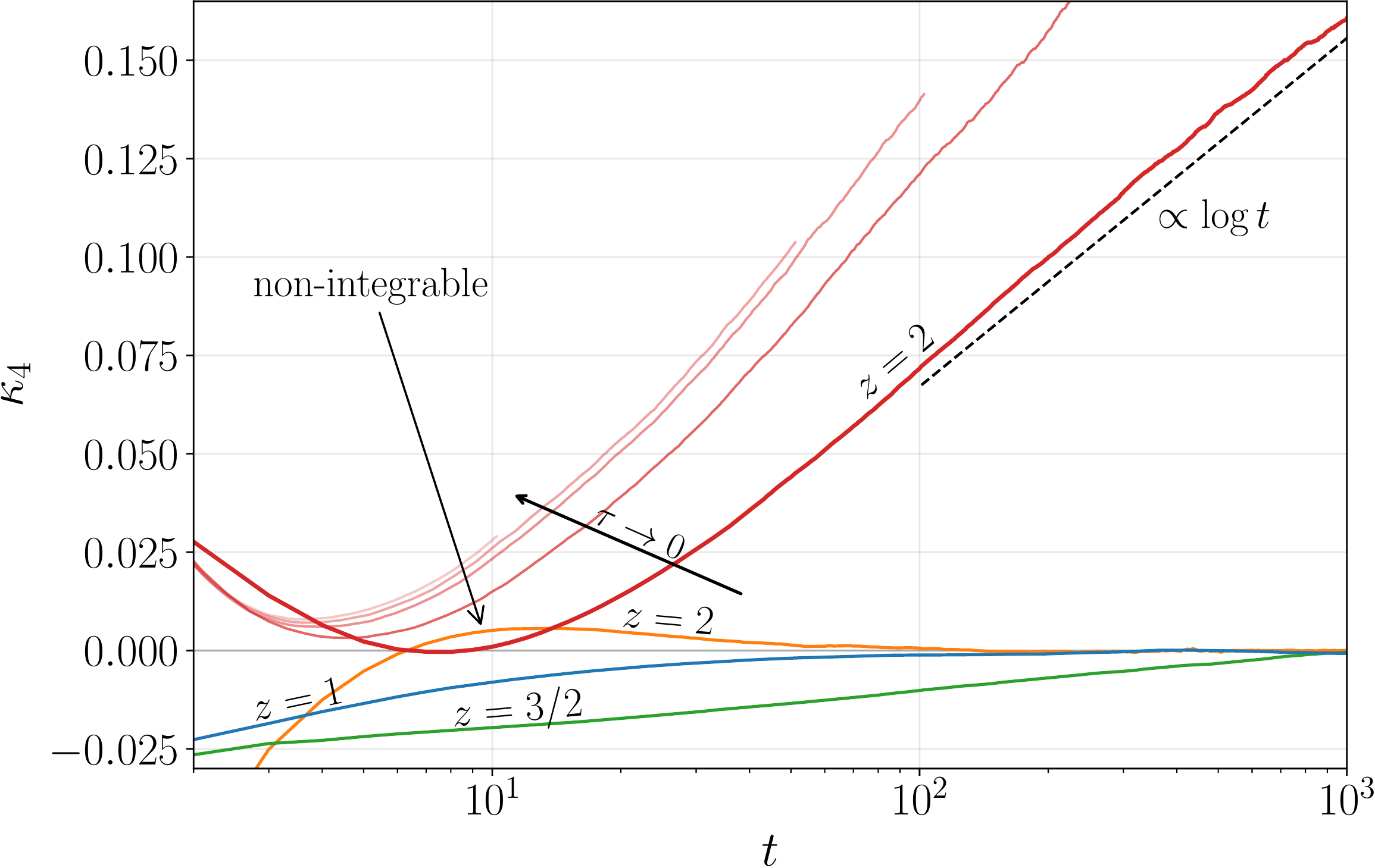}
\caption{Time dependence of $\kappa_{4}(t)$ in different dynamical regimes characterized by the dynamical exponent $z$:
diffusive easy-axis regime (red curves, indicating dependence on time-step $\tau \in \{1, 0.3,0.1,0.05,0.01\}$),
isotropic point (green), ballistic easy-plane regime (blue), and diffusive isotropic non-integrable model (orange). Dashed black line indicates a $\log t$ dependence as a guide to the eye.
Other simulation parameters: timestep $\tau = 1$, length $L = 2^{11}$, anisotropy $\varrho = \gamma = 1$, $N_{\rm avg}=3\times10^{5}$.
The models and conventions are given in \cite{suppmat}.}
\label{fig:figure2}
\end{figure}

{\em Fluctuations in an integrable magnet.}--- We subsequently consider the anisotropic Landau--Lifshitz magnet,
one of the best studied paradigms of interacting spins. In continuous space-time, the model is described by the equation of motion
\begin{equation}
\partial_{t}{\bf S} = {\bf S}\times \partial^{2}_{x}{\bf S} + {\bf S}\times {\mathrm J}\,{\bf S},
\label{eqn:LL}
\end{equation}
with anisotropy tensor $\mathrm{J}={\rm diag}(0,0,\delta)$, representing one of the simplest completely
\emph{integrable} PDEs~\cite{Takhtajan77,Faddeev_book}. Equation~\eqref{eqn:LL} is particularly convenient since
 tuning the anisotropy $\delta$ permits the study of three distinct dynamical regimes \cite{Bojan,KP20,LLMM}:
(i) the `easy-plane' ballistic regime ($z=1$, $\delta<0$), (ii) the easy-axis diffusive regime ($z=2$, $\delta > 0$) and finally (iii)
the isotropic point with superdiffusive spin transport ($z=3/2$, $\delta=0$) -- in exact correspondence with the
dynamical phases of the Heisenberg $XXZ$ spin-$1/2$ chain \cite{superdiffusion_review}.
Indeed, Eq.~\eqref{eqn:LL} is known to be the effective evolution law for the semi-classical eigenstates
(i.e. spin waves of large wavelengths) in the quantum spin chain (see e.g. \cite{NGIV20,MGI21}).

We have performed numerical simulations on the \emph{lattice} counterpart of Eq.~\eqref{eqn:LL} (details in \cite{suppmat}).
One should be aware that a na\"{i}ve lattice discretization of Eq.~\eqref{eqn:LL} does \emph{not} preserve integrability, which may introduce certain spurious effects that affect dynamical properties at large times. This can be overcome by taking advantage of an
exact symplectic integrator based on an integrable regularization in discrete space-time constructed in Ref.~\cite{LLMM}, thereby significantly boosting efficiency of numerical integration (we have verified that the results do not qualitatively change upon varying the time-step $\tau$, see Fig.~\ref{fig:figure2}). The accuracy of our data is only subject to statistical errors due to the size $N_{\rm avg}$ of 
an ensemble of initial conditions sampling the unbiased equilibrium infinite temperature state.

We start by assessing the fluctuations of transferred magnetization
$\mathfrak{J}(t) = \sum_{x>0}(S^3_{x}(t)-S^3_{x}(0))$ by computing the dynamical distribution of the integrated current
$\mathcal{P}(\mathfrak{j}|t)$, rescaled to the timescale of typical fluctuations. There results are collected
in Fig.~\ref{fig:figure1}. Most notably, we observe a significant deviation from Gaussianity in the diffusive case (Fig.~\ref{fig:figure1}c).
In all other regimes of interest, fluctuations appear to be fairly consistent with a Gaussian profile.
To quantify the degree of non-Gaussianity we focus next on the fourth cumulant $\kappa_{4}(t)$, see Fig.~\ref{fig:figure2}, where we 
observe (approximately) logarithmic divergence of $\kappa_{4}$ in the diffusive (i.e. easy-axis) regime,
while in other cases $\kappa_4(t)$ converges to zero.

\begin{figure}[t]
\centering
\includegraphics[width=\columnwidth]{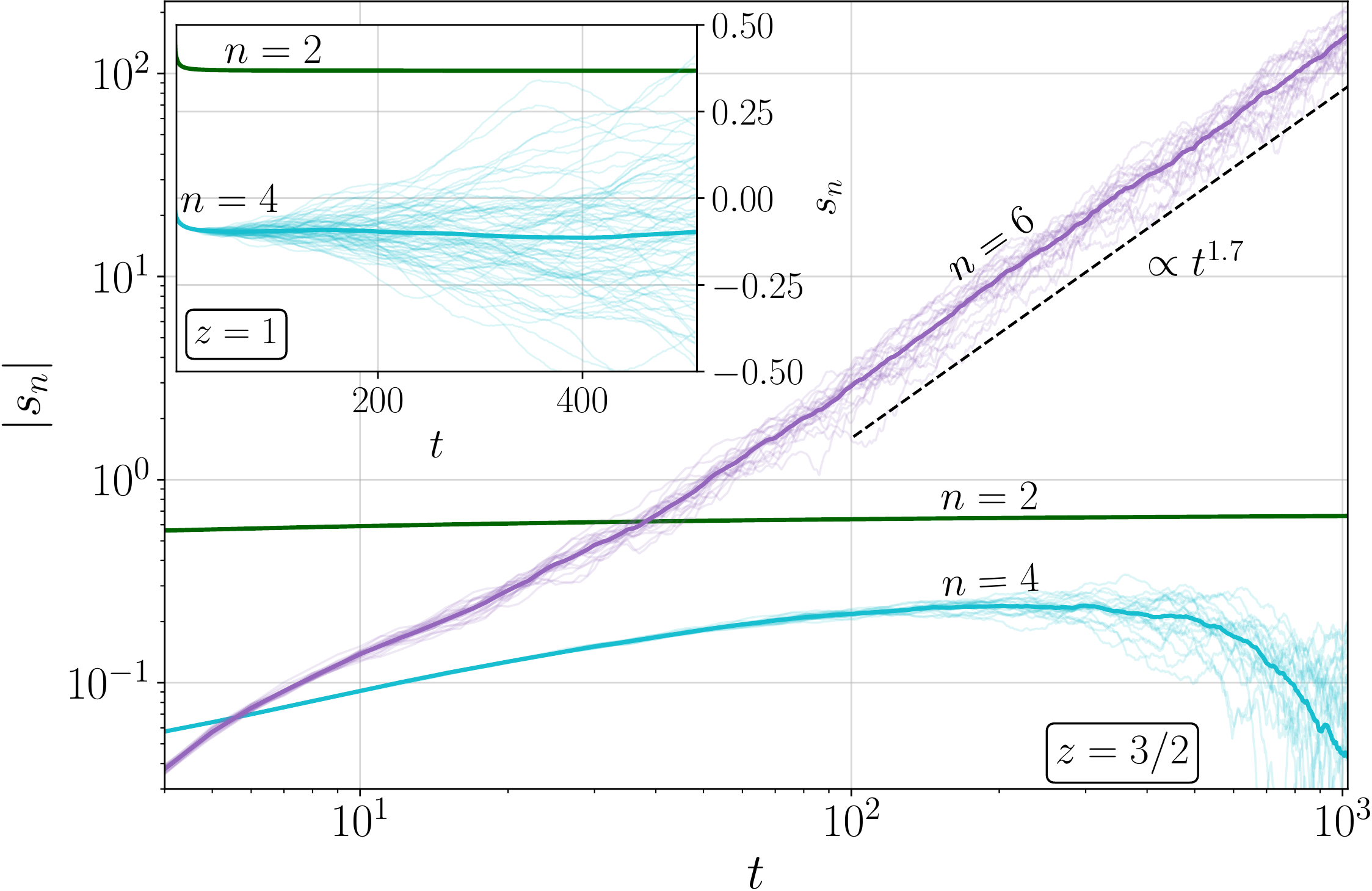}
\caption{Temporal growth of  scaled cumulants $s_{n}(t)$ (absolute values) at (main figure)  the  isotropic point ($z=3/2$) and  (inset) in the easy plane regime ($z=1$). Faint lines show $N_{\rm par} $ partial averages, each over $N_{\rm avg}$ initial random spin configurations. Full line is the total average over $N_{\rm par}\cdot N_{\rm avg}$ initial configurations. Dashed black line indicates the scaling
$s_{6}(t)\sim t^{\nu_{6}}$ with $\nu_{6}\approx 1.7$.
Simulation parameters: timestep $\tau = 1$, length $L = 2^{11}\, (\textrm{main figure})$, $2^{10}\, (\textrm{inset})$, anisotropy $\gamma = 1$ (inset),
$N_{\rm avg}=3\times10^{5}$, $N_{\rm par} = 25\, (\textrm{main figure}), 10^2\, (\textrm{inset})$.
The models and conventions are given in \cite{suppmat}.}
\label{fig:figure3}
\end{figure}

\begin{figure}[b]
\centering
\includegraphics[width=\columnwidth]{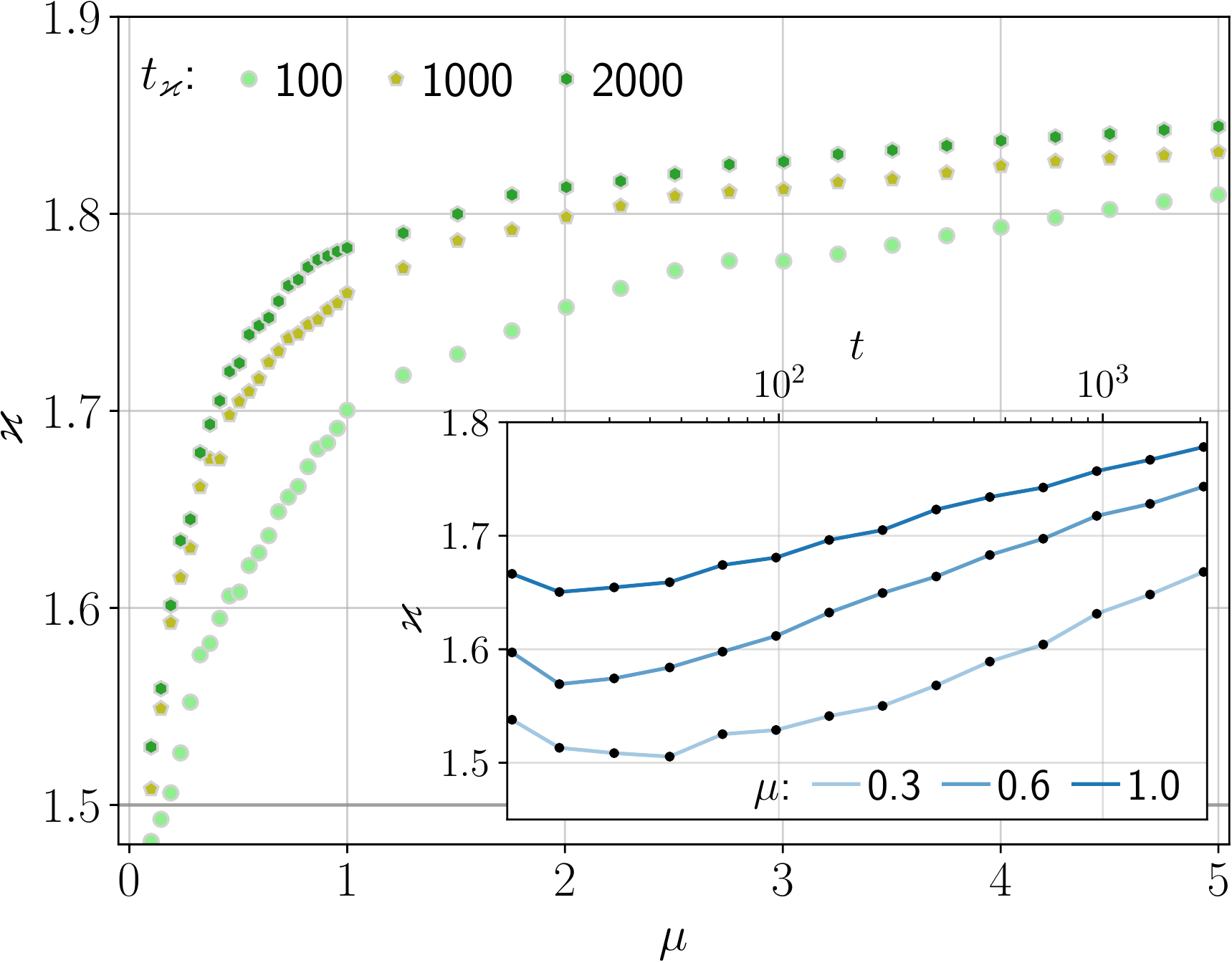}
\caption{Estimated dynamical exponent $\varkappa$ depending on chemical potential $\mu$ for different integration times $t_{\varkappa}$
(with gray line showing $z_{\rm KPZ}=3/2$).  (inset) Time dependnece of the estimated exponent for $\mu \in \{0.3, 0.6, 1.0\}$. Simulation parameters: $\tau = 1$, $L = 2^{12}$, $N_{\rm avg}=$ $3 \times 10^{5}$ (main figure), $10^{6}$ (inset).
The models and conventions are given in \cite{suppmat}.}
\label{fig:figure4}
\end{figure}

{\em Long-time growth of cumulants.---}
The discernible departure from Gaussianity indicates that spin transport in the diffusive phase escapes the usual paradigm of
normal diffusion, as conventionally described within the framework of the MFT.
In the scope of LD theory, another universal feature of stochastic diffusive systems (such as e.g. simple exclusion processes) is the existence of the scaled cumulants 
$s_n(t) = t^{-1/z} \left\langle [\mathfrak{J}(t)]^n\right\rangle^c = ({\rm d}/{\rm d}\lambda)^n F(\lambda|t)|_{\lambda=0}$, where 
$ F(\lambda|t) \equiv t^{-1/z}\log \left\langle e^{\lambda \mathfrak{J}(t)} \right\rangle$.
Following the standard prescription form the literature, the limits of scaled cumulants can be computed from the series expansion of
the scaled cumulant generating function (SCGF) $F(\lambda) = \lim_{\to\infty}F(\lambda|t)$ as
$\lim_{t\to\infty}s_n(t) =  ({\rm d}/{\rm d}\lambda)^n F(\lambda)|_{\lambda=0}$. This scheme however hinges
on certain subtle requirements~\cite{followup} which, as we argue next, may be violated in integrable deterministic dynamics.
%Only provided the family of functions $F(\lambda,t)$ satisfies a set of subtle conditions~\cite{followup}, one can compute the limits of scaled cumulants via Taylor expanding the so called scaled cumulant generating function (SCGF) $F(\lambda) = \lim_{\to\infty}F(\lambda|t)$,
%specifically $\lim_{t\to\infty}s_n(t) =  ({\rm d}/{\rm d}\lambda)^n F(\lambda)|_{\lambda=0}$.
%However, we argue that these conditions may be violated in integrable deterministic dynamics.
Specifically, we show in Fig.~\ref{fig:figure3} that the scaled cumulants diverge in the isotropic and easy-axis regimes of our model, i.e. when $z>1$. At the isotropic point $\delta = 0$ we detect a robust algebraic divergence of the sixth scaled cumulant $s_{6}(t) \sim t^{\nu_6}$
with $\nu_{6} \approx 1.7$, in turn implying divergent $\kappa_{6}(t) = t^{-4/(2z)}\,s_6(t)$. It is worth noting that
such a `higher-order discrepancy' of a tiny amplitude $\kappa_6 \lessapprox 10^{-2}$ on the accessible timescale
(by order of magnitude smaller than in the diffusive regime, cf. Figure~\ref{fig:figure2}) can hardly be discerned
from Figure~\ref{fig:figure1}b), where no noticeable deviations from Gaussianity are visible.
Lastly, in the easy-plane (i.e. ballistic) regime, the scaled cumulant $s_{4}(t)$ converges to a finite value, see inset in Fig.~\ref{fig:figure4}.
Although a reliable extraction of higher scaled cumulants is obstructed by the rapidly growing spread of partial averages,
our data (see \cite{suppmat}) gives no indications of divergent scaled cumulants.
We finally note that upon (strongly) breaking integrability, scaled cumulants are expectedly finite for any value of anisotropy (see \cite{suppmat}).

{\em Fluctuations out of equilibrium.}---A convenient setting that is widely used  for studying fluctuations of charge transfer in one-dimensional
systems away from equilibrium is the two-partition protocol. Several important analytic results have been obtained in this way, predominantly
in the domain of stochastic systems \cite{Johansson2000,TW09,TW09_domain, ER13, BK19, Derrida09}.
To study fluctuations of transferred magnetization, one initializes the system in two semi-infinite partitions in equilibrium states at equal temperatures and opposite chemical potentials $\pm \mu$, related to magnetization densities via $\langle S^{3} \rangle(\mu)=\coth{(\mu)} - \mu^{-1}$. The ensuing dynamical interface region expands asymptotically as $x\sim t^{1/\varkappa}$ while reaching a `local quasi-stationary state'. 
Owing to a finite bias (i.e. a jump in the chemical potential $\mu$), the average integrated current does not vanish and (by assuming algebraic asymptotic scaling) we can accordingly write $\langle \mathfrak{J}(t) \rangle_{\rm noneq} \sim t^{1/\varkappa}$.

Regarding a practical implementation there are now two major stumbling blocks that one has to confront: (I) our simulations reveal (see Fig.~\ref{fig:figure4}) that the running (i.e. time-dependent) algebraic exponent $\varkappa$ converges very slowly towards the expected value at late times, thereby preventing reliable estimation of the stationary distribution $\mathcal{P}(\mathfrak{j})$ of the rescaled current; (II) due to
absence of translational symmetry in the initial state, the sampling size is reduced by a factor of system length $L$ compared to the equilibrium setting. It is nonetheless instructive to expand on point (I). Firstly, we wish to point out that the \emph{nonequlibrium} dynamical
exponent $\varkappa$ should not be \emph{a-priori} identified with the \emph{equilibrium} exponent $z$ that governs the asymptotic growth of the variance (see Eq.~\eqref{z_def}). As we shortly demonstrate, this is delicate matter at the isotropic point ($\delta = 0$) where
the equilibrium dynamical exponent $z_{\rm KPZ}=3/2$ of the Kardar--Parisi--Zhang (KPZ) equation \cite{KPZ} is known to be `protected' by a global
nonabelian symmetry \cite{MatrixModels,superdiffusion_review} that has to be preserved both at the level of the time propagator
and the underlying equilibrium state. Despite the fate of KPZ scaling becomes less obvious upon
departing from equilibrium, a recent experimental study suggests that it might survive \cite{Bloch_KPZ}.

Implementing a two-partition protocol, we numerically extract the running dynamical exponent $\varkappa$ as a function of $\mu$
as shown in Fig.~\ref{fig:figure4}. For any finite simulation time $t_{\varkappa}$, we observe a smooth crossover from $\varkappa \approx 3/2$
in the vicinity of $\mu \to 0$ towards the diffusive exponent $\varkappa \approx 2$ upon approaching strong polarizations $\mu \to \infty$.
This indicates that in spite of a pronounced $\mu$-dependent transient, the running dynamical exponent eventually saturates to $\varkappa=2$.
This analysis is aligned with theoretical expectation: KPZ physics of spin transport is sensitive to explicit breaking of
rotational symmetry (here by the initial nonequilibrium state); our simulations show that the broken symmetry is not dynamically restored, and the dynamics is more reminiscent of the melting magnetic domain at zero temperature \cite{Gamayun_domain,Misguich19}.

{\em Conclusion.}---We numerically investigated the properties
of fluctuations in various dynamical regimes of the one-dimensional lattice Landau--Lifshitz magnet by computing the distribution of the time-integrated spin current
and analyzing the time dependence of its cumulants. Most strikingly, we encountered non-Gaussian typical fluctuations on sub-ballistic scales, comprising both the diffusive easy-axis regime and the isotropic point with superdiffusive spin transport (where the effect is much less pronounced).
 This follows as a consequence of divergent scaled cumulants, which moreover imply that the SCGF  is not a generator of scaled cumulants.
While two-point functions in the easy-axis ($\mathbb{Z}_2$-symmetric) and isotropic (SU(2)-symmetric) regimes have previously been found to excellently match~\cite{Ljubotina17,Ljubotina19,Weiner20,KP20,LLMM}, respectively, the diffusive (Gaussian) and KPZ (Pr\" ahoffer-Spohn~\cite{prahofer}) scaling functions, our new data shows that higher-point functions (or distribution of fluctuations) are distinct from diffusive and KPZ universality classes. 
In particular, as KPZ equation manifestly refers to out-of-equilibrium, breaking detailed balance, the distribution of finite time fluctuations in the KPZ equation are always skewed (eg. Tracy-Widom or Baik-Rains, depending on the initial condition) unlike in our 
equilibrium scenario, where they are symmetric.

 On the ballistic timescale, i.e. at finite magnetization density or in the easy-plane regime at zero magnetization, we found no traces of irregular or non-normal behavior 
(apart from considerably slower convergence of averages compared to a nonintegrable chain).
By explicitly breaking integrability we restored ergodicity and we expectedly recovered both the Gaussian form of typical fluctuations and finite scaled cumulants. This suggests that the observed singularity of the SCGF is subtly linked to the presence of interacting quasiparticles (solitons, see Ref.~\cite{Faddeev_book,AblowitzSegur_book}, whose asymptototic stability is ensured by a hierarchy of
non-trivial conservation laws) which we envision to be responsible for `weak' clustering of temporal multipoint current correlations. This possibility has been discussed in Ref.~\cite{DoyonMyers20}, where it is argued that anomalous fluctuations
could occur on the ballistic scale along the ray corresponding to an \emph{isolated} co-propagating normal mode.
As per \cite{DoyonMyers20} however, a continuous spectrum of normal modes (which is to be anticipated in the Landau--Lifshitz magnet,
by analogy to the quantum Heisenberg chain) need not necessarily be detrimental for Gaussianity.
Concerning the ballistic regime, the absence of any irregularities is thus consistent with the described scenario.
Altough the exact expressions for low-order cumulants have been recently derived \cite {PerfettoDoyon21} by employing the `generalized hydrodynamics' \cite{PhysRevX.6.041065,PhysRevLett.117.207201}, their explicit evaluation crucially relies on the knowledge of the
`flux Jacobian' \cite{Doyon_lectures} which for the particular case of the anisotropic Landau--Lifshitz model is currently out of reach.
The discernible divergence of scaled cumulants that we captured on \emph{subballistic} scales however goes beyond the current capabilities and
for the time being remains entirely elusive. In this view, the most pressing question is to identify a microscopic mechanism responsible for the observed anomalous behavior.

%\begin{table}[t]
%\begin{tcolorbox}[tabulars={@{\extracolsep{\fill}\hspace{5mm}}l|c|c|c@{\hspace{5mm}}},
%fonttitle=\bfseries,fontupper=\normalsize\sffamily,
%boxrule=1pt, adjusted title= fluctuations of transferred magnetization $\mathfrak{J}(t)$, halign title = center, rounded corners = all,
%colback=blue!5!white, colframe=red!50!blue, colbacktitle=blue!20!white, coltitle = black]
%{\tt dynamical exponent} & ${\bf z=1}$     & ${\bf z=3/2}$    & ${\bf z=2}$ \\ \hline\hline
%{\tt typical (CLT)}   & $\tick$ & $\tick$ &  $\textcolor{red}{\bf X}$ \\ \hline
%{\tt rare (LDP)}   & $\tick$ & $\textcolor{red}{\bf X}$ &  $\textcolor{red}{\bf X}$ \\
%\end{tcolorbox}
%\caption{Summarizing the structure typical and rare fluctuation of the net trasferred magnetization (integrated spin current density)
%for various dynamical regimes.}
%\label{table}
%\end{table}

Our hope is that the technical difficulties we encountered in extending our analysis to the nonequilibrium setting can be 
surmounted, as it would help tremendously to establish a more complete phenomenological picture. It would likewise be valuable to complement the
earlier findings of Ref.~\cite{Znidaric_anomalous} by a similar analysis for the case of the anisotropic quantum Heisenberg chain. Our expectation here is that the anomalous structure of dynamical fluctuations in subballistic regimes will also surface at the quantum level. Finally, amidst many recent experimental breakthroughs we firmly believe the time is ripe to initiate a pursuit to find
irregular features in flucuating macroscopic quantities.

\paragraph*{\bf Acknowledgements.}
We thank I. Bloch, S. Gopalakrishnan, V. Pasquier, V. Popkov,  J. Schmidt, J. Zeiher and M. \v{Z}nidari\v{c} for insightful discussions and comments.
ŽK acknowledges support of the Milan Lenarčič foundation.
This work has been supported by the European Research Council (ERC) under the Advanced Grant No.\ 694544 -- OMNES, 
and by the Slovenian Research Agency (ARRS) under the Program P1-0402.

%%%%%%%%%%%%%%%%%%%%%%%%%%%%%%%%%%%%%%%%%
%%%%%%%%%%%%%%%%%%%%%%%%%%%%%%%%%%%%%%%%%

\bibliography{fluctuations}

\clearpage

\onecolumngrid
\begin{center}
\textbf{{\large Supplemental Material for \\ ``Absence of Normal Fluctuations in an Integrable Magnet''}}
\end{center}

\tableofcontents

\section{Integrable discretizations}
In this section, we collect the essential information about the integrable models used in our study.
We employ an integrable space-time discretization of the classical Landau--Lifshitz one-dimensional magnet, whose continuum
theory describe the motion of the classical spin field ${\bf S} = (S^1, S^2, S^3)$ constrained to a two-sphere
${\bf S} \in \mathcal{S}\cong S^{2}$, evolving according to
\begin{equation}
\partial_{t}{\bf S} = {\bf S}\times \partial^{2}_{x}{\bf S} + {\bf S}\times {\mathrm J}\,{\bf S},
\qquad {\rm J} = \textrm{diag} (0, 0, \delta),
\label{sm_eqn:LL}
\end{equation}
where parameter $\delta \in \mathbb{R}$ parametrizes the interaction anisotropy.
Using the $SO(3)$ Lie-Poisson algebra
\begin{equation}
\left \{S^a(x), S^b(y) \right\} = \sum_{c}\varepsilon^{abc}S^c(x) \delta(x-y),
\end{equation}
where $\varepsilon$ is the Levi-Civita tensor, equation \eqref{sm_eqn:LL} can be cast in Hamiltonian form
\begin{equation}
\partial_{t}{\bf S} = \{{\bf S},H\} = -{\bf S}\times \frac{\delta H}{\delta {\bf S}},
\end{equation}
with
\begin{equation}
H = \frac{1}{2} \int \left[\left(\partial_x {\bf S}\right)^2 - {\bf S} \cdot {\rm J} {\bf S}\right] \, \dd x. \label{field_ham}
\end{equation}

Depending on the value of $\delta$, we distiguish between three different dynamical regimes characterized by the dynamical exponent $z$ (in a maximum entropy state with no net magnetization, see Eq. (2) of the main text):
 \begin{alignat}{2}
\delta &\, > 0 \quad  \textrm{easy-axis regime, } &&   z=2, \nonumber\\ 
\delta &\, = 0  \quad \textrm{isotropic regime, }  && z=3/2, \nonumber\\
\delta &\, > 0 \quad   \textrm{easy-plane regime, } && z=1. \label{delta_cases}
\end{alignat} 

\subsection{Space-time discretization}

\begin{figure}[h]
\centering
\begin{tikzpicture}[scale=1.4]
\draw [<->,thick] (-5,1) node (yaxis) [above] {$t$}
        |- (-4,0) node (xaxis) [right] {$\ell$};

\begin{scope}[rotate=45]
\draw[fill=black!5,very thick, drop shadow] (-2,2) rectangle (0,4);
\draw[fill=black!5,very thick, drop shadow] (2,-2) rectangle (4,0);
\draw[fill=black!5,very thick, drop shadow] (0,0) rectangle (2,2);
\draw[fill=black!5,very thick, drop shadow] (2,2) rectangle (4,4);
\draw[fill=black!5,very thick, drop shadow] (0,4) rectangle (2,6);
\draw[fill=black!5,very thick, drop shadow] (4,0) rectangle (6,2);

\draw[fill=black!15,very thick] (2,0) rectangle (4,2);
\draw[fill=black!15,very thick] (0,2) rectangle (2,4);

\draw[dashed] (-1,1) rectangle (1,3);
\draw[dashed] (1,-1) rectangle (3,1);
\draw[dashed] (1,1) rectangle (3,3);
\draw[dashed] (3,-1) rectangle (5,1);
\draw[dashed] (-1,3) rectangle (1,5);
\draw[dashed] (3,1) rectangle (5,3);
\draw[dashed] (1,3) rectangle (3,5);

\foreach \i in {1,...,3} {
	\draw[gray,decoration={markings,mark=at position 0.65 with
    		{\arrow[scale=1.5,>=stealth']{latex}}},postaction={decorate}] (-3+2*\i,5-2*\i) -- (-3+2*\i+1,5-2*\i);
    	\draw[gray,decoration={markings,mark=at position 0.55 with
    		{\arrow[scale=1.5,>=stealth']{latex}}},postaction={decorate}] (-3+2*\i-1,5-2*\i) -- (-3+2*\i,5-2*\i);
	\draw[gray,decoration={markings,mark=at position 0.55 with
    		{\arrow[scale=1.5,>=stealth']{latex}}},postaction={decorate}] (-3+2*\i,5-2*\i-1) -- (-3+2*\i,5-2*\i);
	\draw[gray,decoration={markings,mark=at position 0.6 with
    		{\arrow[scale=1.5,>=stealth']{latex}}},postaction={decorate}] (-3+2*\i,5-2*\i) -- (-3+2*\i,5-2*\i+1);

	\node at (-3+2*\i,5-2*\i) [fill=red!50,inner sep=0pt,minimum size=24pt,draw,rounded corners, drop shadow] (P1i) {$\Phi$};

	\draw[gray,decoration={markings,mark=at position 0.65 with
    		{\arrow[scale=1.5,>=stealth']{latex}}},postaction={decorate}] (-1+2*\i,7-2*\i) -- (-1+2*\i+1,7-2*\i);
    	\draw[gray,decoration={markings,mark=at position 0.55 with
    		{\arrow[scale=1.5,>=stealth']{latex}}},postaction={decorate}] (-1+2*\i-1,7-2*\i) -- (-1+2*\i,7-2*\i);
	\draw[gray,decoration={markings,mark=at position 0.55 with
    		{\arrow[scale=1.5,>=stealth']{latex}}},postaction={decorate}] (-1+2*\i,7-2*\i-1) -- (-1+2*\i,7-2*\i);
	\draw[gray,decoration={markings,mark=at position 0.6 with
    		{\arrow[scale=1.5,>=stealth']{latex}}},postaction={decorate}] (-1+2*\i,7-2*\i) -- (-1+2*\i,7-2*\i+1);

	\node at (-1+2*\i,7-2*\i) [fill=red!50,inner sep=0pt,minimum size=24pt,draw,rounded corners, drop shadow] (P1i) {$\Phi$};
}

\foreach \i in {1,...,2} {

	\draw[gray,decoration={markings,mark=at position 0.65 with
    		{\arrow[scale=1.5,>=stealth']{latex}}},postaction={decorate}] (-1+2*\i,5-2*\i) -- (-1+2*\i+1,5-2*\i);
    	\draw[gray,decoration={markings,mark=at position 0.55 with
    		{\arrow[scale=1.5,>=stealth']{latex}}},postaction={decorate}] (-1+2*\i-1,5-2*\i) -- (-1+2*\i,5-2*\i);
	\draw[gray,decoration={markings,mark=at position 0.55 with
    		{\arrow[scale=1.5,>=stealth']{latex}}},postaction={decorate}] (-1+2*\i,5-2*\i-1) -- (-1+2*\i,5-2*\i);
	\draw[gray,decoration={markings,mark=at position 0.6 with
    		{\arrow[scale=1.5,>=stealth']{latex}}},postaction={decorate}] (-1+2*\i,5-2*\i) -- (-1+2*\i,5-2*\i+1);

	\node at (-1+2*\i,5-2*\i) [fill=red!50,inner sep=0pt,minimum size=24pt,draw,rounded corners, drop shadow] (P1i) {$\Phi$};
}

\foreach \i in {1,...,6} {

	\node at (-3+\i,4-\i) [fill=teal!30,inner sep=0pt,minimum size=24pt,draw,circle,drop shadow] (M1i) {\scriptsize{$\vec{S}^{t}_{\i}$}};
	\node at (-3+\i+1,4-\i+1) [fill=teal!30,inner sep=0pt,minimum size=24pt,draw,circle,drop shadow] (M2i) {\scriptsize{$\vec{S}^{t+1}_{\i}$}};
	\node at (-3+\i+2,4-\i+2) [fill=teal!30,inner sep=0pt,minimum size=24pt,draw,circle,drop shadow] (M3i) {\scriptsize{$\vec{S}^{t+2}_{\i}$}};
	\node at (-3+\i+3,4-\i+3) [fill=teal!30,inner sep=0pt,minimum size=24pt,draw,circle,drop shadow] (M4i) {\scriptsize{$\vec{S}^{t+3}_{\i}$}};
}

\end{scope}

\end{tikzpicture}
\caption{Symplectic circuit from two-body propagators: classical spin degrees of freedom
$\vec{S}^{t}_{\ell}$ (circles) are attach to the middle of the edges of a discrete space-time lattice represented by a tilted checkerboard.
A two-body symplectic map $\Phi_{\tau}$ (red square) that maps a pairs of incoming physical spins forward in time is
attached to the middle of each square plaquette. [The figure has been reproduced from Ref.~\cite{LLMM}.]}
\label{fig:circuit}
\end{figure}
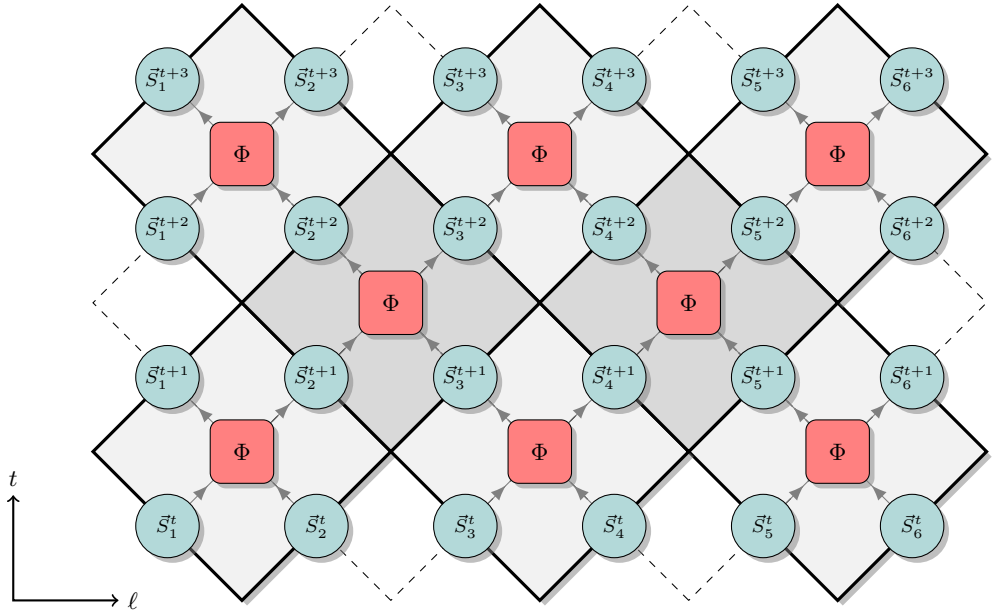

We briefly review an integrable space-time lattice discretization of Eq.~\eqref{sm_eqn:LL} constructed in Refs. \cite{KP20,MatrixModels}.
The main object of the integration scheme is a two-body map
$\Phi_{\tau}: \mathcal{S} \times \mathcal{S} \mapsto \mathcal{S} \times \mathcal{S}$ that provides
a local propagator for a pair of adjacent spins
\begin{equation}
({\bf S}_1, {\bf S}_2) \mapsto  ({\bf S}_1', {\bf S}_2'). \label{two_body}
\end{equation} 
The parameter $\tau \in \mathbb{R}_{+}$ is an ajustable integration timestep.
The two-body propagator \eqref{two_body} serves as the elementary building block of the full (i.e. many-body) propagator $\Phi^{\textrm{full}}: \mathcal{S}^{\times L} \mapsto \mathcal{S}^{\times L}$ over the phase space of $L$ spins (where $L \in 2\mathbb{N}$), consisting
of alternating odd and even steps (see Fig.~\ref{fig:circuit})
\begin{equation}
(\vec{S}^{2t+2}_{2\ell-1},\vec{S}^{2t+2}_{2\ell}) = \Phi(\vec{S}^{2t+1}_{2\ell-1},\vec{S}^{2t+1}_{2\ell}),\qquad
(\vec{S}^{2t+1}_{2\ell},\vec{S}^{2t+1}_{2\ell+1}) = \Phi(\vec{S}^{2t}_{2\ell},\vec{S}^{2t}_{2\ell+1}).
\label{eqn:st}
\end{equation}
By making use of the embedding prescription,
$\Phi^{(j)} = I^{\times (j-1)}\,\times \, \Phi_\tau\, \times\, I^{\times (L-j-1)}$,
%for $j=1,\ldots,L-1$, where $I:\mathcal{S}\mapsto \mathcal{S}$ stands for a local (single-site) unit map $I(\vec{S})\equiv \vec{S}$,
%($\Phi^{(L)}$ correspondingly affects the $L$th and the first spin),
 the full propagator $\Phi^{\rm full}$ for two units of time $t\mapsto t+2$ decomposes as
\begin{equation}
\Phi^{{\rm full}} = \Phi^{\rm even}\circ \Phi^{\rm odd},
\label{full_propagator}
\end{equation}
with odd and even propagators further factorizing into commuting two-body maps,
\begin{equation}
\Phi^{\rm odd} = \prod_{\ell=1}^{L/2} \Phi^{(2\ell)}, \qquad
\Phi^{\rm even}= \prod_{\ell=1}^{L/2} \Phi^{(2\ell-1)}.
\end{equation}
Such an integration scheme constitutes an integrable Trotterization of Eq.~\eqref{sm_eqn:LL}.

\subsection{Integrable two-body propagators}
\subsubsection{Isotropic interaction}
The integrable two-body propagator for the isotropic interaction was obtained in Ref.~\cite{KP20}
\begin{align}
\Phi_{\tau}  (\vec{S}_1, \vec{S}_2) =  \frac{1}{s^2 + \tau^2} \left(s^2\mathbf{S}_1 + \tau^2\vec{S}_2 + \tau \vec{S}_1\times\vec{S}_2, s^2\vec{S}_2 + \tau^2\vec{S}_1 + \tau \vec{S}_2\times\vec{S}_1 \right),  \quad
s^2 = \frac{1}{2} \left(1 + \vec{S}_1 \cdot \vec{S}_2 \right),
\label{discrete_isotropic_eom}
\end{align}
and manifestly preserves the total spin
\begin{equation}
\vec{S}_1 + \vec{S}_2 = \vec{S}_1' + \vec{S}_2'.
\end{equation}

\subsubsection{Anisotropic interaction}

The integrable space-time discretization in the anisotropic case was obtained in Ref.~\cite{LLMM}, which can be compactly expressed
in terms of `Sklyanin' variables $\mathcal K,\, \mathcal{S}^\pm$, satisfying the following Poisson relations,
\begin{equation}
\{\mathcal{K}, \mathcal{S}^{\pm}\} = \mp \ii \varrho \, \mathcal{S}^{\pm} \mathcal{K}, \qquad
\{\mathcal{S}^{+}, \mathcal{S}^{-}\} = -\frac{ \ii \varrho}{2}\left(\mathcal{K}^{2} - \mathcal{K}^{-2}\right),
\label{Sklyanin_bracket}
\end{equation}
with the Casimir function $\mathcal{C}_{0} \equiv \mathcal{S}^{+} \mathcal{S}^{-} + \tfrac{1}{4}\left(\mathcal{K}-\mathcal{K}^{-1}\right)^{2}$.
The two-body propagator takes the form
\begin{align}
\left( \frac{\mathcal{K}'_1}{\mathcal{K}_1} \right)^2 &= 
 \frac{(\nu^2 + \nu^{-2})\left( \mathcal{K}_1 \mathcal{K}_2 +  \mathcal{K}_1^{-1} \mathcal{K}_2^{-1} \right) - 2\mathcal{K}_1 \mathcal{K}_2^{-1} - (w^2 + w^{-2}) \mathcal{K}_1^{-1} \mathcal{K}_2 + 4w \mathcal{S}^+_1 \mathcal{S}^-_2 + 4w^{-1}\mathcal{S}^-_1 \mathcal{S}^+_2}
{(\nu^2 + \nu^{-2})\left( \mathcal{K}_1 \mathcal{K}_2 +  \mathcal{K}_1^{-1} \mathcal{K}_2^{-1} \right) - 2\mathcal{K}_1^{-1} \mathcal{K}_2 - (w^2 + w^{-2}) \mathcal{K}_1 \mathcal{K}_2^{-1} + 4w \mathcal{S}^-_1 \mathcal{S}^+_2 + 4w^{-1}\mathcal{S}^+_1 \mathcal{S}^-_2},  \nonumber\\
\left( \frac{\mathcal{K}'_2}{\mathcal{K}_2} \right)^2 &= 
\frac{(\nu^2 + \nu^{-2})\left( \mathcal{K}_1 \mathcal{K}_2 +  \mathcal{K}_1^{-1} \mathcal{K}_2^{-1} \right) - 2\mathcal{K}_1^{-1} \mathcal{K}_2 - (w^2 + w^{-2}) \mathcal{K}_1 \mathcal{K}_2^{-1} + 4w \mathcal{S}^-_1 \mathcal{S}^+_2 + 4w^{-1}\mathcal{S}^+_1 \mathcal{S}^-_2}
 {(\nu^2 + \nu^{-2})\left( \mathcal{K}_1 \mathcal{K}_2 +  \mathcal{K}_1^{-1} \mathcal{K}_2^{-1} \right) - 2\mathcal{K}_1 \mathcal{K}_2^{-1} - (w^2 + w^{-2}) \mathcal{K}_1^{-1} \mathcal{K}_2 + 4w \mathcal{S}^+_1 \mathcal{S}^-_2 + 4w^{-1}\mathcal{S}^-_1 \mathcal{S}^+_2}, \nonumber \\
\begin{bmatrix}
(\mathcal{S}^{\pm}_1)^{\prime} \\
(\mathcal{S}^{\pm}_2)^{\prime}
\end{bmatrix}
&= \Omega(w^{\pm 1})
\begin{bmatrix}
\mathcal{S}^{\pm}_{1} \\
\mathcal{S}^{\pm}_{2}
\end{bmatrix},
\label{Sklyanin_eom}
\end{align}
with
\begin{equation}
\Omega(w) = \frac{1}{\mathcal{K}_1 \mathcal{K}_2 w - (\mathcal{K}_1 \mathcal{K}_2 w)^{-1}} \!
\begin{bmatrix}
\mathcal{K}^{\prime}_{1}\mathcal{K}_{2} - (\mathcal{K}^{\prime}_{1}\mathcal{K}_{2})^{-1} & \kern-1em
(\mathcal{K}^{\prime}_{1}/\mathcal{K}_{1})w -(\mathcal{K}_{1}/\mathcal{K}^{\prime}_{1})w^{-1} \\
(\mathcal{K}^{\prime}_{2}/\mathcal{K}_{2})w -(\mathcal{K}_{2}/\mathcal{K}^{\prime}_{2})w^{-1} & \kern-1em
\mathcal{K}^{\prime}_{1}\mathcal{K}_{2} - (\mathcal{K}^{\prime}_{1}\mathcal{K}_{2})^{-1}
\end{bmatrix},
\end{equation}
and parameters
\begin{equation}
\nu = e^{\varrho}, \qquad w = e^{\ii \varrho \tau}.
\end{equation}
The anisotropy of interaction is controlled by a parameter $\varrho$ with the following domains:
\begin{flalign}
\textrm{--} &\textrm{ the \emph{easy-axis} regime with anisotropy } \varrho \in \mathbb{R}_+, && \nonumber\\
\textrm{--} &\textrm{ the \emph{easy-plane} regime with anisotropy } \gamma \in [-\pi/2, \pi/2]. && \label{rho_list}
\end{flalign}
The easy plane anisotropy $\gamma$ is related to the easy axis anisotropy via analytic continuation as $\varrho \to \ii \gamma$. 
Sklyanin's spins are bijectively related to classical spins via
\begin{equation}
\mathcal{K} = e^{\varrho \, S^{3}}, \qquad
%\mathcal{S}^{3} = \sinh{(\varrho \, S^{3})},\qquad
\mathcal{S}^{\pm} =  F_{\varrho}(S^{3})S^{\pm}, \quad F_\varrho(s) \equiv \sqrt{\frac{\sinh^{2}(\varrho) - \sinh^2(\varrho s)}{1-s^2}}.
\label{Sklyanin_vars}
\end{equation}
The propagator \eqref{Sklyanin_eom} preserves the third component of magnetization
\begin{equation}
S^3_1 + S^3_2 =  (S^3_1)' + (S^3_2)'. 
\end{equation}

\subsection{Discrete Noether current}
The full propagator has a global $U(1)$ symmetry which implies a conserved charge $Q = \sum_{\ell} q_\ell$, representing the total magnetization
along the third axis, $q_\ell = S^3_\ell$. Due to the even-odd structure of the dynamics, Eq.~\eqref{full_propagator},
the local density satisfies a pair of discrete continuity equations
\begin{equation}
\frac{1}{\tau}\left(q_{2\ell}^{2t+2} - q_{2\ell}^{2t}\right) + j_{2\ell+1}^{2t+1} - j_{2\ell}^{2t} = 0,  \qquad
\frac{1}{\tau}\left( q_{2\ell+1}^{2t+2}  - q_{2\ell+1}^{2t} \right) + j_{2\ell+2}^{2t} - j_{2\ell+1}^{2t+1} = 0,
\label{discrete_continuity}
\end{equation}
which are satisfied by a current of the form
\begin{equation}
j_{\ell}^{t} \equiv j(\ell,t) = \frac{1}{\tau}\big(q_\ell^{t+1} - q_\ell^t \big).
\label{discrete_j}
\end{equation}

\subsection{Ensemble average}
We introduce the stationary maximum entropy measure $\rho_1$ on $\mathcal{S}$ (i.e. Gibbs states at infinite temperature)
and define a separable measure $\rho_L$ by extending it to the many-body phase $\mathcal{M}_{L}\equiv \mathcal{S}^{\times L}$, namely
\begin{equation}
\rho_{L} = \prod_{\ell=1}^{L}\rho_{1}(\vec{S}_{\ell}).
\label{many_body_measure}
\end{equation}
For the normalized one-body measure we consider the grand-canonical ensemble with a chemical potential $\mu \in \mathbb{R}$,
\begin{equation}
\rho_{1}(\vec{S}) = \frac{1}{4\pi} \frac{\mu}{\sinh \mu} e^{\mu \, S^3}.
\end{equation}
Passing to the thermodynamic limit by sending $L\to \infty$, the grand-canonical partition function per site is given by
\begin{equation}
\mathcal{Z}_{1}(\mu)=\int_{\mathcal{S}}\dd \Omega\,\rho_{1}(\vec{S}) = \frac{4\pi \sinh{\mu}}{\mu},
\end{equation} with $\dd \Omega$ denoting the volume element on $\mathcal S$.
Writting $\dd \Omega^{\rm full} \equiv \prod_{\ell=1}^{L} \dd \Omega(\vec{S}_{\ell})$, the
average of a local observable $\mathcal{O}$ is computed as
\begin{equation}
\langle \mathcal{O} \rangle_{\mu} = \lim_{L\to \infty}\big[\mathcal{Z}^{L}_{1}(\mu)\big]^{-1}
\int_{\mathcal{M}_{L}}\dd \Omega^{\rm full}\,\rho_{L}\,\mathcal{O}.
\end{equation}
The average magnetization is related to $\mu$ via
\begin{equation}
\langle S^{3} \rangle(\mu) = \coth{\mu}-1/\mu.
\end{equation}

\subsection{Field theory limit}

The reduction to the field theory involves two steps. Firstly, we send $\tau \to 0$ to obtain a Hamiltonian lattice model
\begin{equation}
H_{\rm LLL}^{\varrho} \simeq \sum_{\ell=1}^{L} \log \left[ (\sinh \varrho)^2\, \mathcal{S}^{0}_{\ell}\mathcal{S}^{0}_{\ell+1} + \mathcal{S}^{1}_{\ell}\mathcal{S}^{1}_{\ell+1} + \mathcal{S}^{2}_{\ell} \mathcal{S}^{2}_{\ell+1} + (\cosh \varrho)^2\, \mathcal{S}^{3}_{\ell}\mathcal{S}^{3}_{\ell+1}\right],
\end{equation}
in terms of Sklyanin spins
$
\mathcal{S}^{0}\equiv \frac{1}{2}(\mathcal{K}+\mathcal{K}^{-1}),\enspace
\mathcal{S}^{1}\equiv \frac{1}{2}(\mathcal{S}^{+} + \mathcal{S}^{-}),\enspace 
\mathcal{S}^{2}\equiv \frac{1}{2\ii}(\mathcal{S}^{+} - \mathcal{S}^{-}),\enspace
\mathcal{S}^{3}\equiv \frac{1}{2}(\mathcal{K}-\mathcal{K}^{-1}).
$
 We then reinstate a lattice spacing $\varDelta$ and subsequently retain smooth lattice spin configurations
$\vec{S}_\ell = \vec{S}(x=\Delta \ell)$, $\vec{S}_{\ell \pm 1} = \vec{S}(x) \pm \varDelta \partial_{x} \vec{S} + \frac{1}{2}\varDelta^{2} \, \partial_{x}^{2} \vec{S} + \mathcal{O}(\varDelta^{3})$ by keeping only variations at the leading non-trivial order $\mathcal{O}(\varDelta^{2})$.
Finally taking the continuum limit $\varDelta \to 0$, both the Sklyanin variables and the anisotropy parameter 
have to be rescaled by the lattice spacing, namely
\begin{equation}
\mathcal{S}^{0} = 1 + \mathcal{O}(\varDelta^{2}), \qquad
\mathcal{S}^a = \varDelta \, \vec{S}^{a} + \mathcal{O}(\varDelta^{2}) \quad {\rm for}\quad a=1,2,3,
\end{equation}
and $\varrho \rightarrow \varDelta \, \varrho$, thus recovering the Hamiltonian of the Landau--Lifshitz field theory,
\begin{equation}
H = \frac{1}{2} \int \left[\left(\partial_x {\bf S}\right)^2 - {\bf S} \cdot {\rm J} {\bf S}\right] \, \dd x, \qquad
{\rm J} = \textrm{diag} (0, 0, \varrho^2), \label{field_ham}
\end{equation}
upon making the identification $\delta = \varrho^2$.

\subsection{Integrable easy-plane regime: higher scaled cumulants}
\begin{figure}[h]
\includegraphics[width=0.8\linewidth]{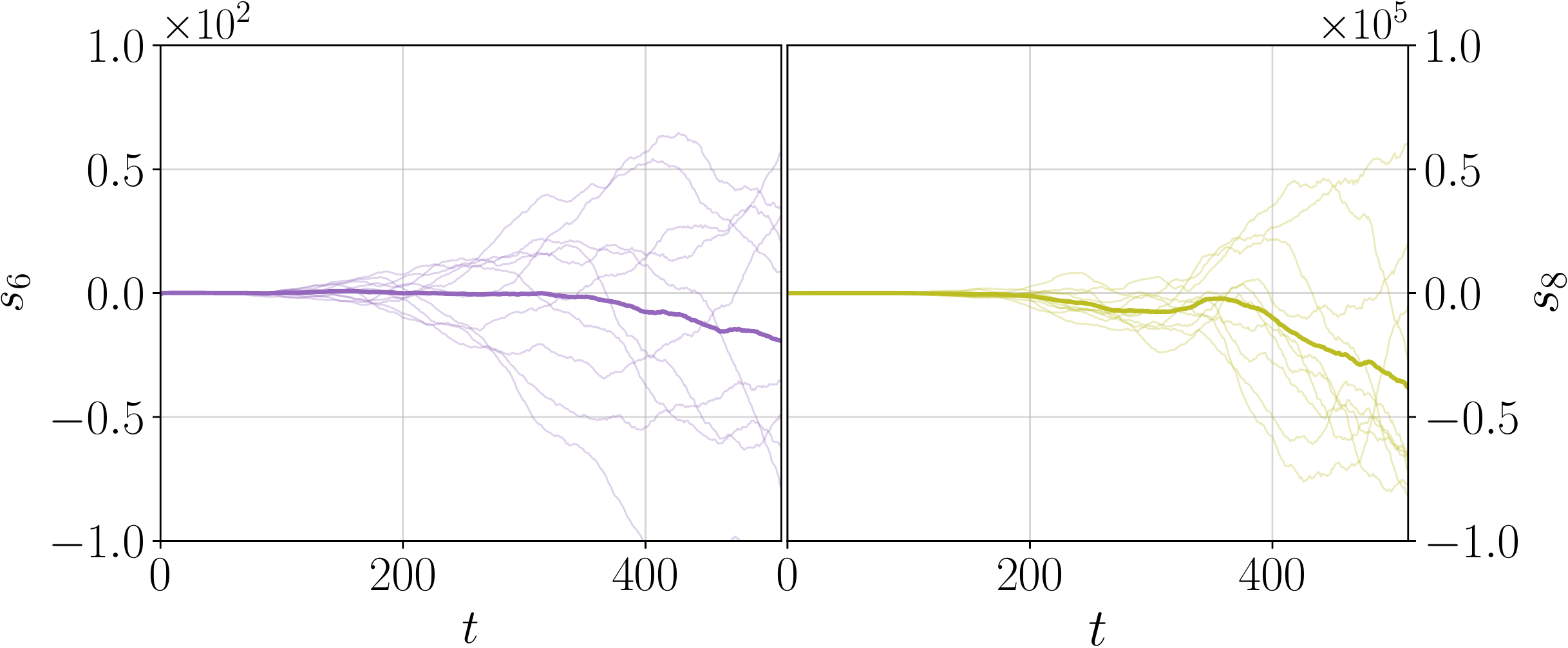}
\caption{Temporal growth of scaled cumulants $s_{6}, s_8$ in the easy-plane regime ($z=1$). Faint lines show $N_{\rm par} $ partial averages, each over $N_{\rm avg}$ initial random spin configurations. Full line is the total average over $N_{\rm par}\cdot N_{\rm avg}$ initial configurations.
%Shaded area shows the time-dependent statistical uncertainty of the average estimated as $\sigma(t) = \sigma^{\textrm{par}}(t)/\sqrt{N_{\textrm{par}}}$, where $\sigma^{\textrm{par}}(t)$ is the standard deviation of partial averages.
Simulation parameters: timestep $\tau = 1$, length $L = 2^{10}$, anisotropy $\gamma = 1$,
$N_{\rm avg}=6\times10^{6}$, $N_{\rm par} = 10$.}
\label{fig:SM_figure1}
\end{figure}

Here we provide the numerical data on scaled higher cumulants $s_n(t)$ associated to the time-integrated current density (refer to the main text for definitions) in the ballistic ($z=1$) easy-plane regime of \eqref{Sklyanin_eom}.
The non-zero scaled cumulants $n=6$ and $n=8$ are shown in Fig.~\ref{fig:SM_figure1}.
%While averages for $s_2$ and $s_4$ appear to stabilize at finite values, 
The rapidly growing spread of partial averages makes it difficult to make any definite statements about the asymptotic values of $s_{n\geq 6}(t)$. We nevertheless observe no significant divergent behaviour in any of the cumulants and attribute the late-time
behaviour of $s_{6}(t)$ and $s_8(t)$ to statistical fluctuations (thin lines) which are an order of magnitude larger than the magnitude of the average (thick line). We emphasize that the full average is not the arithmetic average of partial averages owing to the nonlinearity of the cumulants.

\section{Non-integrable discretizations}
It is instructive to make a comparisson to an ergodic Hamiltonian dynamics. To this end, we also consider the scaled cumulants of the time integrated current density in nonintegrable discretizations of the anisotropic Landau-Lifhsitz field theory, Eq.~\eqref{sm_eqn:LL},
as outlined below.

\begin{figure}[h]
\includegraphics[width=0.6\linewidth]{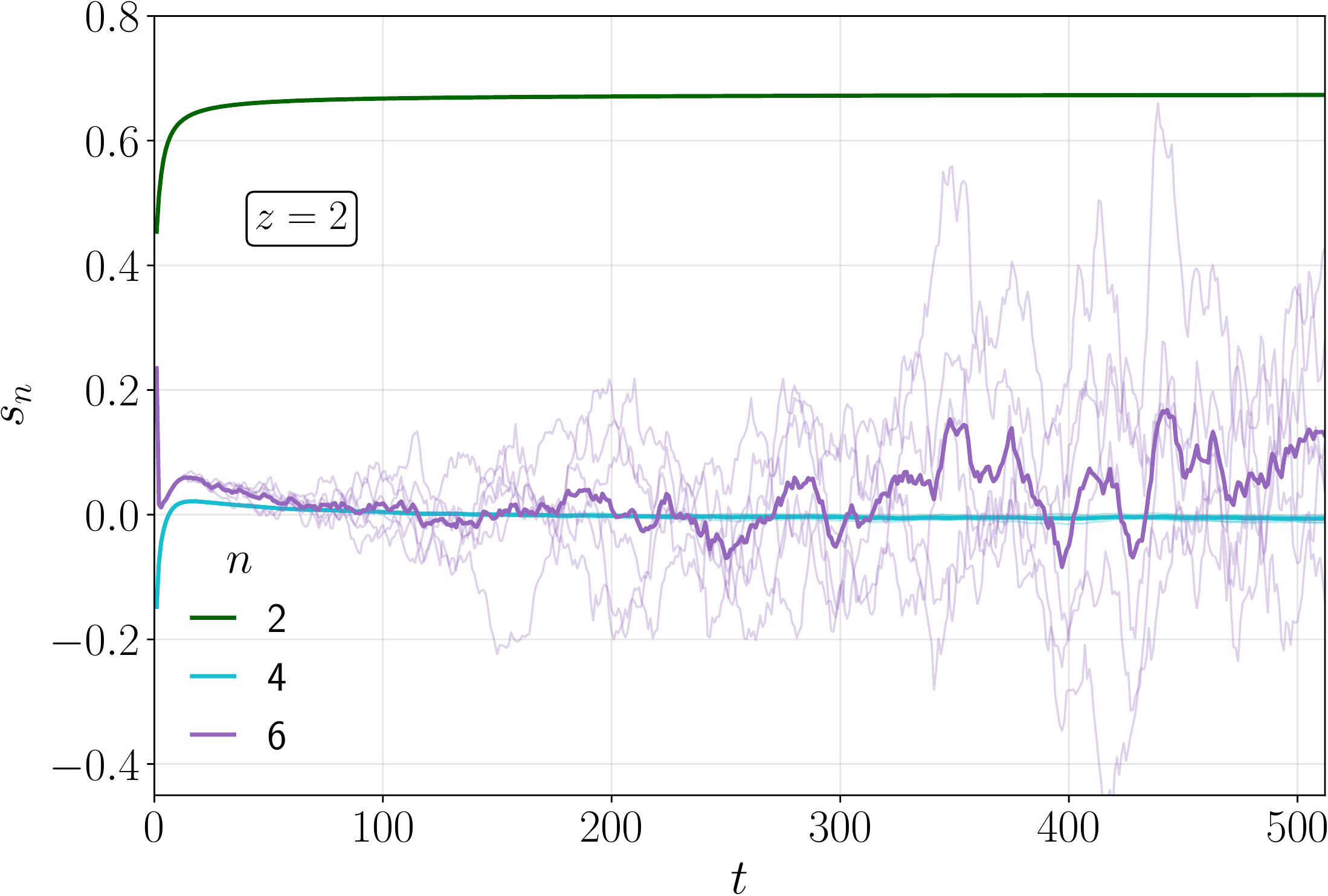}
\caption{Temporal growth of scaled cumulants $s_n$ in the nonintegrable isotropic regime (z=2). Faint lines show $N_{\rm par} $ partial averages, each over $N_{\rm avg}$ initial random spin configurations. Full line is the total average over $N_{\rm par}\cdot N_{\rm avg}$ initial configurations.
%Shaded area shows the time-dependent statistical uncertainty of the average estimated as $\sigma(t) = \sigma^{\textrm{par}}(t)/\sqrt{N_{\textrm{par}}}$, where $\sigma^{\textrm{par}}(t)$ is the standard deviation of partial averages.
Simulation parameters: timestep $\tau = 1$, length $L = 2^{10}$, $N_{\rm avg}=3\times10^{6}$, $N_{\rm par} = 6$.}
\label{fig:SM_figure2}
\end{figure}

\subsection{Two-body propagator for isotropic interaction}
The isotropic dynamics can be efficiently discretized by employing the same Trotterization scheme (cf. Eq.~\eqref{eqn:st}) as in the integrable case, only replacing the two-body propagator $\Phi_\tau$. Deriving an non-integrable two-body map is straighforward and does not require any integrability techniques. Instead, it suffices to find the explicit solution to the two-body problem governed by the simple Hamiltonian
\begin{equation}
H =  \vec{S}_1 \cdot \vec{S}_{2},
\end{equation} 
yielding the equations of motion (see also the Supplemental Material of Ref.~\cite{undular})
\begin{equation}
\dot{\vec{S}}_{1} = \vec{S}_{1} \times \vec{S}_{2}, \qquad \dot{\vec{S}}_{2} = \vec{S}_{2} \times \vec{S}_{1}.
\end{equation}
A non-integrable two-body propagator $\Phi_\tau$ corresponds to evaluating the solution of the above at time $t = \tau$, yielding
\begin{equation}
S_{\ell}' = e^{\ii \tau S} S_{\ell}  e^{-\ii \tau S},  \qquad {\rm with}\quad
S = S_1 + S_2, \quad S_{\ell} = \vec{S}_{\ell} \cdot \pmb{\sigma},\quad \ell = 1, 2, \label{nint_iso}
\end{equation}
where $\pmb{\sigma} = (\sigma^{1}, \sigma^{2}, \sigma^{3})$ is a vector of Pauli matrices.
The discrete dynamics given  by Eq.~\eqref{nint_iso} permits efficient simulation of non-integrable spin dynamics with local isotropic
spin interaction.

\subsection{Scaled cumulants}

Simulating the non-integrable isotropic model, cf. Eq.~\eqref{nint_iso}, we expectedly observe diffusive spreading at late time, characterized
by dynamical exponent $z=2$. We moreover computed the scaled cumulants $s_n(t)$ (shown in Fig.~\ref{fig:SM_figure2}). We find that $s_4$ rapidly decays towards zero. The large spread of partial averages of $s_6(t)$ does not allow for an extraction of its asymptotic value, but we observe no systematic divergent behaviour (cf. Fig.~3 of the main text), indicating the existence of scaled cumulants in the nonintegrable model (\ref{nint_iso}). Moreover $s_6$ asyptoticaly decaying towards zero  is within the range of statistical fluctuations of the data.

\end{document}